\documentclass[10pt]{article}

\usepackage{graphicx}

\usepackage{graphicx,psfrag}

\def \v {\vskip 0.2cm }

\def \s {\sigma}

\def \i {{\hbox {i}}}

\def \cd {{\cal D}}

\def \vep {\varepsilon}

\def \a {\alpha}

\def \b {\beta}

\def \T {{\hbox {Tr}}}

\def \g {\gamma}

\def \G {\Gamma}

\def \t {\tilde}

\def \h {\hat}

\def \E {{\bf E}}

\def \i {{\hbox{ i}}}

\def \vep {\varepsilon}

\def \D {\Delta}

\def \d {\delta}

\def \n {\noindent}

\begin{document}
\title{Asymptotic Behavior of Partition Functions with Graph Laplacian }

\author{O. Khorunzhiy}

%\date{}

\maketitle

\begin{abstract}

We introduce the matrix sums that represent a discrete analog of the matrix integrals of random matrix theory. The summation runs over the set $\G_n$ of all possible 
$n$-vertex graphs $\g_n$ weighted by $\exp\{ - \beta\ \T \D_n\}, \beta >0$, where  
$\D_n= \D(\g_n)$ is the analog of the  Laplace operator determined on $\g_n$. Corresponding probability measure on $\G_n$ reproduces the well-known Erd\H os-R\'enyi ensemble of random graphs. Here it plays the same role as that played by the Gaussian Unitary Invariant Ensemble (GUE) in matrix models.

Regarding an analog of the matrix models with quartic potential, we study the cumulant expansion of related partition functions. We develop a diagram technique and describe the combinatorial structure of the coefficients of this expansion in two different asymptotic regimes $\beta = O(1)$ and
$\beta = O(\log n)$ as $n\to\infty$.

\end{abstract}

\noindent 
{\it Keywords:} Laplace operator on graph, partition function, Erd\H os-R\' enyi random graphs, connected diagrams, Catalan numbers, P\'olya equation.

\section{Introduction}
During last three decades, the studies of matrix models of theoretical physics 
have deeply influenced a number of branches of modern mathematics and mathematical physics. The central notion here is the partition function 
$Z_N(\b,  Q)$ given by the integral over the set ${\cal M}_N$ 
of all Hermitian $N$-dimensional matrices $H$
$$
Z_N(\beta, Q) = \int_{{\cal M}_N} \exp\{ - \beta \T H^2 + Q(H)\} d H
= C_N \E_{\hbox{\tiny GUE}}^{(\b)} \{e^{Q(H)}\},
\eqno (1.1)
$$
where  $Q$ is a "potential" function, $d H$ is the Lebesgue 
measure, 
\mbox{$C_N= Z_N(\b,0) $} is the normalizing constant, and $ \E_{\hbox{\tiny GUE}}^{(\b)}
\{\cdot \}$ denotes the mathematical expectation with respect 
to the probability measure ${\cal P}_N$ with the density 
$$
C_N^{-1} \exp\{ - \b \T H^2\} = C_N^{-1} \exp\{ - \b \sum_{i,j=1}^N \vert H_{ij}\vert ^2\}.
\eqno (1.2)
$$
The measure ${\cal P}_N$ supported on ${\cal M}_N$ 
generates the Gaussian Unitary Invariant Ensemble of random matrices abbreviated by 
GUE (see monograph \cite{M} for the detailed description of the ensemble and its properties).

The first non-trivial example of (1.1) is given by the quartic potential
$$
Q(H) = {g\over N} \T H^4.
\eqno (1.3)
$$
The matrix model (1.1), (1.3) has served as the source
of a series of deep results establishing connections between 
orthogonal polynomials, integrable systems, moduli spaces of curves and such combinatorial structures as  maps (see \cite{BIZ} and \cite{EM} for  the earlier and more recent results and references
and \cite{D} for the review). One of the simplest but important result is 
that the leading term of the formal asymptotic expansion of variable
$$
 {1\over N^2} \log  \E_{\hbox{\tiny GUE}}^{(\b)} \left\{\exp\left({g\over N } \T H^4 \right) \right\}
\eqno (1.4)
$$
in the limit $N\to\infty$ is given by the series in powers of $g$ with the coefficients determined by  the numbers 
of $4$-valent two-vertex maps dual to the famous quadrangulations 
of the compact Riemann manifold (see \cite{Z} for introductory description and references therein).

It should be noted that the GUE and its real symmetric and symplectic analogs represent very special class of random matrices. It is natural to ask about matrix models when the mathematical expectation in (1.4)
is taken with respect to a measure different from that determined by GUE.

In present paper we introduce a discrete analog of  (1.1), where
the integration over  ${\cal M}_N$ is replaced by 
the sum over the set $\G_n$
of  all possible simple graphs $\gamma_n$ with $n$ vertices. 
In this setting the "kinetic energy" term $\T H^2$ is replaced by 
$\T \D_n$, where $\D_n=\D(\gamma_n)$ is the discrete analog of the Laplace operator
determined on the graph $\gamma_n$. Corresponding Gibbs weight
$\exp\{- \beta \ \T \D(\g_n)\}, \b >0$ generates the probability measure 
$\mu_n(\b)$ on $\G_n$.

A simple but non-trivial property of this ensemble plays a very  important role in what follows.
The observation is that the probability space $(\G_n, \mu_n)$   coincides with the 
widely known Erd\H os-R\'enyi ensemble of random graphs with $n$ vertices (see e.g. \cite{B}).
In this ensemble, the indicator functions of edges are represented by jointly  
independent Bernoulli random variables. 
As far as we know, this connection between Erd\H os-R\'enyi  random graphs 
and the Gibbs measure $C_n \exp\{- \beta \ \T \D(\g_n)\}$ was not observed before.

Our aim is  to consider the asymptotic behavior
of $F_n = \log \E_{\mu_n} \left\{ \exp( Q_n )\right\}$ where $Q_n$ is given by the analog
of quartic potential (1.3) and to explore the combinatorial structures that arise in this problem. We develop a diagram technique to study the cumulant expansion of $F_n$. 
We show that the leading terms of this expansion are related with the number
of connected diagrams on the set of two-valent vertices. 
We derive recurrent relations for the numbers of such diagrams and describe the coefficients
of the cumulant expansions of $F_n$ in two different asymptotic regimes 
when $\b = O(1)$ and $\b = O(\log n)$ as $n\to \infty$. 
These recurrent relations generalise those for the Catalan numbers. Corresponding  generating function
verifies an equation similar to the P\'olya equation for the generating function of the rooted Cayley trees.

\section{Graph Laplacian, matrix sums and random graphs}

Given  a finite graph with the set $V_n = \{v_1,\dots , v_n\}$ of labelled vertices and 
$E_m= \{e^{(1)}, \dots , e^{(m)}\}$ the set of simple non-oriented edges, the discrete analog of the Laplace operator $\D(\g)$ on graph $\g$  can be determined as (see for example, \cite{Mo}) by relation
$$
\D(\gamma) = \partial ^* \partial,
\eqno (2.1)
$$
where $\partial $ is the difference operator determined on the space of complex functions on vertices $V_n\to {\bf C}$ and $\partial^*$ is its conjugate determined on the space of complex functions on edges $E_m\to {\bf C}$.

It can be easily shown that in the canonical basis, the linear  operator 
$\D(\gamma)=\D_n $ has 
$n\times n$ matrix
with the elements 
$$
\D_{ij} = \cases{ \deg(v_j) ,& if $i=j$, \cr
-1, & if $i\neq j$ and $(v_i,v_j)\in E$,\cr
0, & otherwise,\cr
}
\eqno (2.2)
$$
where $\deg(v)$ is the vertex degree. 
If one considers the $n\times n$ adjacency matrix $A = A(\g)$ of the graph $\g$, 
$$
A_{ij} = \cases{ 1 , & if $(v_i,v_j)\in E$, $i\neq j$,\cr
0, & otherwise,\cr}
\eqno (2.3)
$$
then one can rewrite the definition of  $\D$ (2.2) in the form 
$$
\D_{ij} = B_{ij} - A_{ij}, \quad {\hbox {where}} \ \ B_{ij} = \delta_{ij} \sum_{l=1}^n A_{il}, 
\eqno (2.4)
$$
where $\delta$ is the Kronecker symbol
$$
\delta_{ij} = \cases{1, & if $i=j$, \cr
0, & if $i\neq j$.}
$$
It follows from (2.1) that $\D(\gamma_n)$ has positive eigenvalues.

Let us consider the set $\G_n$ of all possible simple non-oriented graphs $\g_n$ with the set $V=V_n $ of $n$ labelled vertices. Obviously, $\vert \Gamma_n\vert = 2^{n(n-1)/2}$.
Given an element $\gamma\in \G_n$, it is natural to consider the trace $\T \D(\gamma)$ as 
the  total energy of the graph $\gamma$. Then we can 
assign to each graph $\gamma_n$ the Gibbs weight
$\exp\{ - \b\  \T \D(\g_n) \}$, $\beta >0$ and  introduce the partition function
$$
Z_n(\b,Q) = \sum_{\g_n \in \G_n } \exp\{ - \b\  \T \D_n + Q(\g_n)\},
\eqno (2.5)
$$
where $\D_n = \D(\gamma_n)$ and $Q$ is an application $\Gamma_n \to {\bf R}$ that we specify later. 
Using the fact that 
$\T \D = \T \partial^*\partial$, one can consider (2.5) as a discrete analog of the partition function
(1.1). We will see that this is especially interesting in the case of quartic potential (1.3). 

Let us note that  we should normalize the sum  (2.5) by 
$\vert \G_n\vert$, 
but 
this does not play any role with respect to our results. In what follows, we omit subscript $n$ in $\D_n$.

Definition (2.4) implies that
$$
\T \D = \sum_{i=1}^n \D_{ii} = \sum_{i,j=1}^n A_{ij} = 2\sum_{1\le i< j\le n} A_{ij}.
\eqno (2.6)
$$
Then we can rewrite (2.5) in the form
$$
Z_n(\b,Q) = 
\sum_{\g_n \in \G_n } e^{  Q(\g_n)} \prod_{1\le i<j\le n} e^{-2\beta A_{ij}}.
\eqno (2.7) 
$$
It is easy to see that 
$$
Z_n(\b,0) =  \left( 1 + e^{-2\b}\right)^{n(n-1)/2}.
\eqno (2.8)
$$
Then the normalized partition function can be represented as
$$
\h Z_n(\b,Q) = Z_n(\b,Q)/Z_n(\b,0) = \E_{\b} \left\{ e^{Q(\gamma)}\right\},
\eqno (2.9)
$$
where $\E_{\b}\{\cdot \}$ denotes the mathematical expectation
with respect to the measure  supported on the set $\G_n$.
This measure assigns to each element $\gamma \in \G_n$ the probability
$$
P_n(\gamma) = {\displaystyle e^{-2\beta \vert E(\gamma)\vert}\over 
 \left( 1 + e^{-2\b}\right)^{n(n-1)/2}},
 $$
 where $E(\gamma) $ denotes the set of  edges of  the graph $\gamma$.

Given a couple $(x,y)$, $x,y \in \{1,\dots,n\}$, one can determine a random variable
$a_{xy}$ 
on the probability space $(\G_n,P_n)$ that is the indicator function of the edge $(v_x,v_y)$
$$
a_{xy} (\gamma)= \cases{ 1, & if $(v_x,v_y)\in E(\gamma)$,\cr 
0, & otherwise.\cr}
$$
It is easy to show that   the random variables $\{ a_{xy}, 1\le x< y\le n\}$ are jointly independent and are of the same Bernoulli distribution depending on $\beta$ such that
$$
a_{xy}^{(\b)} = \cases{ 1 ,& with probability $ {\displaystyle e^{-2\beta}\over 
\displaystyle 1+ e^{-2\beta}}=p$,
\cr
0, & with probability $1-p$.\cr }
\eqno (2.10)
$$

The probability space $(\G_n, P_n)$ is known as the Erd\H os-R\'enyi (or Bernoulli) ensemble of random graphs with the edge probability $p$. Since the series of pioneering
papers by Erd\H os and R\'enyi, 
the asymptotic properties of graphs  $(\G_n, P_n)$, 
such as the size and the number of 
connected components, the maximal and minimal vertex degree and many others, are extensively studied (see \cite{B,JLR}).
Spectral properties of corresponding random matrices $A$ (2.3) and $\D$ 
(2.4) are considered in a series of  papers (in particular, see \cite{BG,E,KKM,KSV,KS,MF,RB}).
In present paper we  study the random graph ensemble 
$(\G_n, P_n)$ from another point of view motivated by the 
asymptotic behavior of partition functions (2.9) with certain "potentials" $Q$.
 
 \section{Partition functions and diagram technique}

In the previous section, we have shown that the Gibbs weight $\exp\{ - \beta\ \T \D(\gamma)\}$ generates the probability measure on graphs  equivalent to that determined by the  Erd\H os-R\'enyi  ensemble of random graphs.
This gives us an important tool for direct  computation of averages of the form (2.9). Another important point is that this Gibbs weight leads us to the correct definition of  discrete analogs  of the matrix models
(1.1) and in particular of the matrix models with quartic potentials (1.4). That is why  one cannot  
neglect the Laplacian form of the Gibbs measure and start simply with computations of averages with respect to $P_n$.

\subsection{Analog of the quartic potential}

Let us determine the discrete analog of the partition function (1.1) with quartic potential (1.3). 
Once $\T H^2$ replaced by $\T (\partial^*\partial) = \T \D$, 
it is natural to consider 
$$
\T (\partial^*\partial \partial^*\partial) = \T \D^2
$$ 
as the analog of $\T H^4$. Then the partition function 
(2.5) reads as
$$
Z_n(\beta, g) =
\sum_{\g_n \in \G_n } \exp\{ - \b \  \T \D_n + g_n\T \D^2\},
\eqno (3.1)
$$
where $g_n$ is to be specified. It follows from (2.3) and (2.4) that 
$$
\T \D^2 = \T B^2 + \T A^2= \sum_{i,j=1}^n (A^2)_{ij} + \sum_{i,j=1}^n A_{ij}.
$$
Then, using (2.6) and repeating computations of (2.7) and (2.8), we obtain representation
$$
\h Z_n(\b, g )= Z_n(\b, g )/ Z_n(\b, 0 ) = \left( { 1 + e^{-2\b'}\over 
1+ e^{-2\b}} \right)^{n(n-1)/2} \ \E_{\b'} \{ e^{g_nX_n}\},
\eqno (3.2) 
$$
In this relation, we have 
denoted $\b' = \b - g_n$ and introduced the random variable
$$
X_n = \sum_{i,j,l=1}^n a_{il} a_{lj},
\eqno (3.3)
$$
where  $a_{ij}$ are jointly independent random variables of  the law (2.10) with 
$\beta $ replaced by $\beta'$. 
The average $\E_{\b'}$ denotes the corresponding mathematical expectation.
In what follows, we omit the subscripts $\beta$ and $\beta'$
when they are not necessary.
We study the limiting behavior of the cumulants of the random variable $X_n$
in two different asymptotic regimes. The first limiting transition  is determined by the choice
$\b_n = O(\log n)$ as $n\to\infty$. We study this case in Section 4. In Section 5,
we consider the second asymptotic regime given by  of $\beta_n = const$.

\subsection{Cumulant expansion}

Using the fact that $a_{ij} $ are bounded by $1$ and $X_n\le n^3$, we can write  that
$$
\log \E \{e^{gX_n}\} = \sum_{k=1}^\infty {g^k\over k!} Cum_k(X_n),
\eqno (3.4)
$$
where $Cum_k(X_n)$ is the $k$-th cumulant of random variable $X_n$.
$$
Cum_k(X_n) = {d^k\over dg^k}\left( \log \E\{ e^{gX_n}\}\right) \vert_{g=0}.
\eqno (3.5)
$$
Denoting by $Y_\alpha$ the random variable $a_{il}a_{lj}$ with 
triplet $\alpha = (i,l,j)$,  we can write that
$$
Cum_k(X_n) = \sum_{\{\alpha_1,\dots, \alpha_k\}} 
cum\{Y_{\alpha_1}, \dots, Y_{\alpha_k}\},
\eqno(3.6)
$$
where the sum runs over all possible values of $\alpha_s, s=1,\dots,k$ and
$$
cum\{Y_{\alpha_1}, \dots, Y_{\alpha_k}\}= {d^k\over
dz_1\cdots dz_k} \log \E\{\exp(z_1Y_{\alpha_1}+
\dots +z_k Y_{\alpha_k})\}\vert_{z_r=0}.
\eqno(3.7)
$$
The variable $cum\{Y_{\alpha_1}, \dots, Y_{\alpha_k}\}$ is also known in probability theory
as the semi-invariant of $k$ random variables $Y$.

Let us introduce a graphical representation of the set of variables $Y$.
Given $k$ values $\alpha_1, \dots,\alpha_k$, we represent the set of  random variables
$Y_{\alpha_1}, \dots,Y_{\alpha_k} $ by the set of $k$ labelled vertices
$U=(u_1,\dots,u_k)$. Regarding two variables $Y_{\alpha_r}$ and
$Y_{\alpha_s}$, we join corresponding vertices $u_r$ and $u_s$ by an edge $\vep(r,s)$ if and only if 
 $Y_{\alpha_r}$ and 
$Y_{\alpha_s}$ have at least one variable $a$ in common. 
Considering 
all possible  couples $(r,s), 1\le r<s\le k$ and 
drawing corresponding edges, we obtain a graph that we denote by 
$G_k =(U_k, {\cal E}_k)$.  It is clear that this graph $G_k = G_k(\vec \alpha_k)$
depends on particular value of the variable 
$\tilde \alpha_k = (\alpha_1,\dots,\alpha_k)$.
The following proposition is a well-known fact from
the probability theory of random fields \cite{MM}.

\v
\n {\bf Lemma 3.1.} 
 {\it The semiinvariant $cum\{Y_{\alpha_1}, \dots, Y_{\alpha_k}\}$ is not equal to zero
if and only if the graph $G_k(\tilde \alpha_k)$ is connected. }
\v
\n {\it Proof.}
Let us first note that if two random variables $Y_{\alpha_r}$
and $Y_{\alpha_s}$  have no variables $a$ in common, then they are  independent. 
Clearly, the same observation is true for the subfamilies of random variables $Y$.
This means that 
if the graph $G_k(\tilde \alpha_k)$
consists of two or more non-connected components, then the corresponding subsets of random variables
$Y$ are jointly independent. The characteristic property of the semiinvariants is that it vanishes 
in this case \cite{MM}. This   completes the proof of Lemma 3.1. 

\v
Let us note that the fact that  $Y_{\alpha_r}$ and 
$Y_{\alpha_s}$ have one or more  variables $a$ in common means that corresponding 
variables $i,l,j$ coincide. In particular, the fact that 
$Y_{\alpha_r}$ and 
$Y_{\alpha_s}$ have exactly one   variable $a$ in common implies 
that one of the eight possibilities for the sets $(i_r,l_r,j_r)$ and $(i_s,l_s,j_s)$ occurs. For example,
this happens when
$$
(i_r=i_s, l_r=l_s, j_r\neq j_s)\quad {\hbox{or}}\quad  (i_r=l_s, l_r=i_s, j_r\neq j_s).
\eqno (3.8)
$$
We will refer to such cases as to the direct and inverse gluing, respectively.

\v
\n {\bf Lemma 3.2.} 
{\it Given $\beta$ and $g$ fixed, the number of terms in  $Cum_k(X_n)$ is given by relation}
$$
\#\{Cum_k(X_n)\} = O(n^{k+2})\quad as \ n\to\infty.
\eqno (3.9)
$$
\v 
\v

\n {\it Proof.}  To draw an edge $\vep(r,s)$ of the graph $G_k$ means to make
equal at least two variables $a$ taken from $Y_{\alpha_r}$ and $Y_{\alpha_s}$, respectively. 
To make equal  two or more $a$'s means to make equal some of 
the variables $(i_r,l_r,j_r)$ 
and $(i_s,l_s,j_s)$.  

Let us describe the process of drawing edges of $G_k$ step by step.
One starts with  $3k$ variables $i,j,l$ that can take values $1,\dots,n$ independently.
Each gluing of two variables $a$ diminishes by $2$ the number of variables that move independently.
To make the graph connected, we have to draw at least $k-1$ edges, so  
we are forced to perform 
at least $k-1$ gluings. When this is done,  the number of variables that move independently
is less or equal to $3k - 2(k-1) = k+2$. Maximizing the number of variables that can take different values, 
from $1$ to $n$ independently, we obtain 
$$
n(n-1)\cdots (n-(k-1)+1) = O(n^{k+2})
$$
 terms.
Lemma 3.2 is proved.

\subsection{Connected diagrams}
 
Let us further develop the graphical representation of the set of variables $Y$ and
give more details for the description of corresponding
connected graphs $\t G_k$. 
At this stage we make no difference between the direct and inverse gluings (3.8).

First let us note that 
each variable $Y_\alpha$ by itself can be represented  by a graph of three vertices corresponding
to variables $i,l$ and $j$ joined by two edges
denoting random variables $a_{il}$ and $a_{lj}$, respectively. 
Slightly modifying this, we can say that $Y_{\alpha_r}$ is represented by a vertex 
$u_r$ with two off-spreads representing variables 
$a_{i_rl_r}$ to the left and $a_{l_rj_r}$ to the right from $u_r$. We say that the vertices $u_r$ are
two-valent.
In $\t G_k$, there are $k$ such two-valent vertices  joined by $k-1$ edges that we refer now to as to arcs.
The arcs join two different off-spreads that we glue during the procedure described in the proof of the Lemma 3.2. 
The family of $k$ two-valent vertices together with $k-1$ arcs represent {\it a diagram} that we denote
by $\d_k$. This diagram provides more information than the graph $\t G_k$ because it shows 
exactly which off-spreads are glued between themselves. If one forgets the two-valent 
structure of $u_r$'s,
one gets the tree $\t G_k$.  One of the possible example of $\d_k$ and corresponding $\t G_k$ for $k=5$ is given on figure 1.

%%%%%%%%%%%%%%%%%%%%%%%%%%%%%%%%%%%%%%%%%%%%%%%%

\begin{figure}[htbp]
\centerline{\includegraphics[width=12cm]{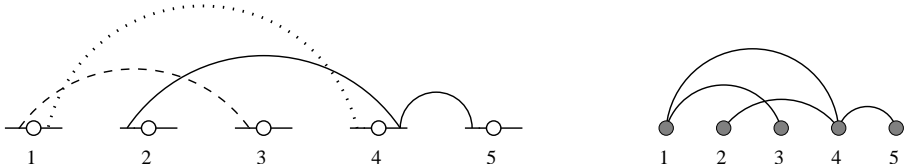}}
\caption{\footnotesize{Connected diagram $\d_k$ and corresponding tree $\tilde G_k$ for $k=5$.}}
\end{figure}

%%%%%%%%%%%%%%%%%%%%%%%%%%%%%%%%%%%%%%%%%

One can see that several off-spreads can be glued together. In this case we say that they are colored
by the same color. We call the  off-spreads that remain non-glued as the free off-spreads and leave them non-colored (or grey).
On figure 1 the diagram $\d_5$ contains three grey elements and  three color groups of $2$, $2$, and $3$ elements. 
   
Clearly, one can draw several diagrams that represent the same coloring of the off-spreads.
An example of such two diagrams $\delta_k$ and $\delta_k'$ with $k=5$  is given on figure 2.
We call the diagram {\it the reduced diagram} when the off-spreads of the same color
are joined by the arcs connecting the nearest neighbors. On figure 2 the diagram $\d_k$ is the reduced one,
the diagram $\d_k'$ is not. 

%%%%%%%%%%%%%%%%%%%%%%%%%%%%%%%%%%%%%%%%%%%%%%%%

\begin{figure}[htbp]
\centerline{\includegraphics[width=12cm]{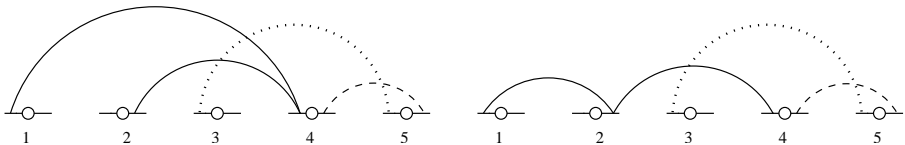}}
\caption{\footnotesize{Non-reduced diagram $\delta_{k}'$ and the equivalent reduced one.}}
\end{figure}

%%%%%%%%%%%%%%%%%%%%%%%%%%%%%%%%%%%%%%%%%

This argument explains the difference between the diagrams we have and the set of trees on labelled vertices.
Indeed, the diagrams $\d_k$ and $\d_k'$ are equivalent but the corresponding trees are not.

With this diagram representation,  we see that the leading contribution to (3.9) comes
from the family $\cd_k$ of connected acyclic reduced diagrams $\d_k$   drawn on the set 
of $k$ two-valent vertices with ordered off-spreads. 
The number of such diagrams $d_k = \vert \cd_k\vert $ 
is determined in the next section.
We complete this section with  the following simple proposition.

\v
{\bf Lemma 3.3.}
{\it Let us consider a  connected reduced diagram $\d_k$ that have $r$ color groups of arcs,
with $\mu_1, \mu_2, \dots, \mu_r$ arcs in each color group. Then 
$$
\mu_1 + \mu_2 +\dots + \mu_r = k-1.
\eqno (3.10)
$$
There are $k-r+1$ grey elements in $\delta_k$.
}

\v
{\it Proof. } 
Each diagram generates a tree on $k$ vertices $u_1, \dots, u_k$. So the total number of arcs
is equal to $k-1$. The group of $\mu_s$ arcs of the same color produces 
$\mu_s+1$ color elements. The total number of colored elements is
$k-1+r$. Then the number of non-colored (grey) elements in $\d_k$ is $2k - (k-1+r) = k-r+1$.

\section{Sparse random graphs}

In present section we study the case when the "temperature " $T_n= 1/\beta_n$  vanishes 
when $n \to\infty$
$$
\beta_n = {1\over 2} \log \left({n\over c}\right) (1+o(1)),\quad n\to\infty
\eqno (4.1a)
$$
with some  $c>0$. 
This corresponds to the random graph ensemble (3.2) with vanishing edge probability
$$
p_n =
\left( {\bar c\over n}\right)^{(1+o(1))}, \quad n\to\infty,
\eqno (4.1b)
$$
where we denoted $\bar c = ce^{2g(n,c)}$ and $g(n, c) = g/c$. In this paper we consider the limiting transition when $n,c\to \infty$ and $c=o(n)$. The case of $n\to\infty, c= const$ will be studied in separate publication. 
The main results of this section are presented by the following two statements.

\v 
\n {\bf Theorem 4.1.}

\n {\it Given $k \ge 2$, there exists the limit
$$
\lim_{n, \bar c \to\infty, \bar c = o(n)}{1\over n\bar c^{k+1}} Cum_k(X_n) = 2^{k-1}  d_k ,
\eqno (4.2)
$$
where  the numbers 
$d_k, k\ge 2$ are determined by following recurrent relation
$$
d_k =  2k d_{k-1} + \sum_{j=1}^{k-2} {k-1 \choose j} (j+1)(k-j)\  d_{j}\,  d_{k-1-j}\ , \ k \ge 3
\eqno (4.3)
$$ 
with the initial condition $d_2=4$}.

\v 

Let us consider the  auxiliary numbers 
$$
h_k = {(k+1) d_k\over k!}
\quad{\hbox{for}}\ \  k\ge 2.
$$
It is easy to deduce from (4.3) that the sequence $h$ can be determined by the  following recurrent relation
$$
h_k = {k+1\over k} \sum_{j=0}^{k-1} h_j \ h_{k-1-j}, \quad h_0 =1.
\eqno (4.4)
$$

\n {\bf Proposition 4.2.}
 
\n {\it The generating function
$
h(z) = \sum_{k=0}^\infty  {h_k} z^k
$
is determined in the complex domain $R_{1/8}= \{z\in {\bf C}: \vert z\vert < 1/8\}$
and verifies there the following equation
$$
h(z) = \exp\{2z h(z)\}.
\eqno (4.5)
$$
It follows from (4.5) that }
$$
h_k = 2^{k}{(k+1)^{k-1}\over k!}, \quad k\ge 1.
\eqno (4.6)
$$
\v

Using (4.4), it is easy to find the  first values of $d$ 
$$
d_2= 4, \ d_3 = 32, \ d_4 = 400, \ d_5 = 6912, \ d_6 = 153664.
$$
The general form of $\{d_k\}$ follows from (4.6)
$$
d_k = 2^k (k+1)^{k-2}, \quad k\ge 2.
\eqno (4.7)
$$

Equation (4.5) is similar to the P\'olya equation for the generating function
of the rooted trees on labelled vertices \cite{O,R} (see also relation (4.17) below). 
The explicit expression (4.6) 
resembles  the number of Cayley trees on $k+1$ vertices. 
However, we did not find any obvious one-to-one correspondence between
the set of Cayley trees on $k+1$ vertices and the $k$-vertex diagrams we count. 
While the skeletons of our diagrams  are given by trees, 
the maximal degree of these trees is bounded by $4$. Also,
there are equivalent diagrams that have different tree skeletons.
This makes the set of the diagrams we study quite different from the
family of Cayley trees. Relation (4.4) generalizes recurrent relations for the Catalan numbers. 
Up to our knowledge, the class of diagrams ${\cal D}_k$ 
as well as the numbers $d_k$ were not considered previously.

\v

\n {\it Proof of  Theorem 4.1.}
Let us separate the set of all possible values of triplets $\a_1,\dots, \a_k$
containing $n^{3k}$ elements   into the classes of equivalence labelled by
 $\d_k\in {\cal D }_k$. This is done in obvious way. Then we can write that 
$$
\sum_{\a_1, \dots ,\a_k} cum\{Y_{\a_1}, \dots, Y_{\a_k}\} =
\sum_{\d_k\in \cd_k}   {\cal N}(\d_k)\  cum \{\bar Y(\d_k)\} (1+o(1)),\quad n\to\infty
\eqno (4.8)
$$
where ${\cal N}(\d_k)$ denotes  the number of elements in the equivalence class labelled by $\d_k$,
$$
cum \{\bar Y(\d_k)\}= cum\{Y_{\bar a(\d_k)_1}, \dots, Y_{\bar a(\d_k)_k}\},
$$
and $\bar a(\d_k)$ is one of the representative of this equivalence class. 
For instant, we can choose   $\bar a(\d_k)= (\a_1, \dots, \a_k)$ 
with minimal possible values of $i_1$, $l_1$, $j_1$, $i_2$, and so on, in the way  such that 
$(\a_1, \dots, \a_k)$ belongs to the class $\d_k$. 
It follows from  Lemma 3.2 that 
$$
{\cal N}(\d_k) = n(n-1)\cdots (n-k-1) = n^{k+2}(1+o(1)), \quad n\to\infty.
\eqno (4.9)
$$
Using  the basic property  of the semi-invariants, we can write that \cite{MM}
$$
cum \{\bar Y(\d_k)\} = \sum_{\pi_k} \E\{\tilde Y_{T_1(\pi_k;\bar \a(\d_k))}\}\cdots
\E\{\tilde Y_{T_\s(\pi_k;\bar \a(\d_k))}\} (-1)^{\s-1} (\s-1)!\ ,
\eqno (4.10)
$$
where the sum runs over all unordered partitions $T_1,\dots,T_\s$ of the set
of vertices $U_k=\{u_1,\dots, u_k\}$; that is over all families of  non-empty non-intersecting subsets 
$T_s$ of $U_k$ giving in sum all $U_k$. The second argument of $T$'s reminds us that 
the variables $\bar \a(\delta_{k})_s$ attached to $u_s$ are chosen according to the rule
prescribed by $\d_k$. Random variable  $\tilde Y_{T_s(\pi_k;\bar \a(\d_k))}$ is given by the 
product of corresponding random variables $Y$.

Let us consider the term of the sum (4.10) that corresponds to the trivial partition
 $\pi^{(1)}_k$ of $U_k$ 
consisting of one subset only: $T_1=U_k$. This term is given by the 
average value 
$$
\E \{Y_{\bar a(\d_k)_1} \cdots Y_{\bar a(\d_k)_k}\} = W(\pi^{(1)}_k).
$$
According to Lemma 3.3, there are $k-r+1$ non-colored elements in diagram $\d_k$,
where $r$ is the number of color groups of arcs in $\d_k$. 
These grey elements represent independent random variables $a$ that are also independent
from elements belonging to color groups.
Then we obtain  the factor 
$$
(\E a)^{k-r+1} = \left({\bar c\over n}\right)^{k-r+1}.
$$
  Each color group
produces the factor 
$
\E a^{\mu_s+1} = \E a = \bar c/n
$ 
and there are $r$ color groups.
Then the weight of the partition $\pi_k^{(1)}$ is 
$$
W(\pi^{(1)}_k)= \left({\bar c\over n}\right)^{k+1}.
\eqno (4.11)
$$
Now it is clear that any other partition $\pi_k$ produces the term
of the order $o((\bar c/n)^{k+1}) $ that is evidently smaller than that of (4.10). 
More precisely, we can write that
$$
\E\{\tilde Y_{T_1(\pi_k;\bar \a(\d_k))}\}\cdots
\E\{\tilde Y_{T_\s(\pi_k;\bar \a(\d_k))}\} =
\left({\bar c\over n}\right)^{k+1+\chi},
\eqno (4.12)
$$
where $\chi$ is equal to the number of the arcs $\theta$ of $\d_k$ such that
the left and right feet of $\theta$ belong to different subsets $T'$ and $T''$
of the partition $\pi_k$ under consideration. We say that these arcs are
cut by partition $\pi_k$. 
In particular, the partition $\pi_k^{(k)}$ of $U_k$ into $k$ subsets
produces the factor
$$
W(\pi^{(k)}_k)= \left({\bar c\over n}\right)^{2k}
$$
because all $k-1$ arcs of $\d_k$ are cut by this partition.

Remembering that each arc can be drawn in the direct and inverse sense (3.8) and 
gathering formulas (3.6), (4.8), (4.9) and (4.11), we conclude that
$$
{1\over n \bar c^{k+1}}  Cum_k(X_n) = 2^{k-1}d_k (1+o(1)), \quad 1\ll \bar c \ll n,
$$
where  $d_k = \vert \cd_k\vert$ is the total number of diagrams $\d_k$. 
Let us  derive recurrent relations for $d_k$.

Let us consider
one particular diagram $\delta_k$ and denote by $P_{j}(\delta_{k})$ some part of $\d_k$
that consists of certain $j$ vertices and all arcs that join them. 

Let us now consider the last vertex  $u_k$. The following two cases are possible:

(a) there is only one arc that joins $u_k$ with $P_{k-1}(\d_k)$ and 

(b) there are two arcs that join $u_k$ with $P_{k-1}(\d_k)$.

\noindent
Clearly, the case when $u_k$ is joined with $P_{k-1}(\d_k)$ by three or more
arcs is prohibited
because of the  absence of cycles in corresponding tree $\tilde G_k$.

In the first case $P_{k-1}(\delta_{k})$ is a connex diagram. 
One can join this diagram with $u_k$ by an arc that have the left  foot supported 
either on grey element or on the maximal element of the color group. If there are
$r$ colour groups in $P_{k-1}(\delta_{k})$, then there are $k-r$ grey elements
(see Lemma 3.3). There are $r$ maximal color elements. Thus one can choose $k$ elements to put the left foot of the arc.
The right foot can be put on one of the two elements of $u_k$. 
Clearly, $\vert P_{k-1} \vert = d_{k-1}$. So, the case (a)
produces $2kd_{k-1} $ connected diagrams.
On figure 3 we illustrate this situation.

%%%%%%%%%%%%%%%%%%%%%%%%%%%%%%%%%%%%%%%%%%%%%%%%

\begin{figure}[htbp]
\centerline{\includegraphics[width=8cm]{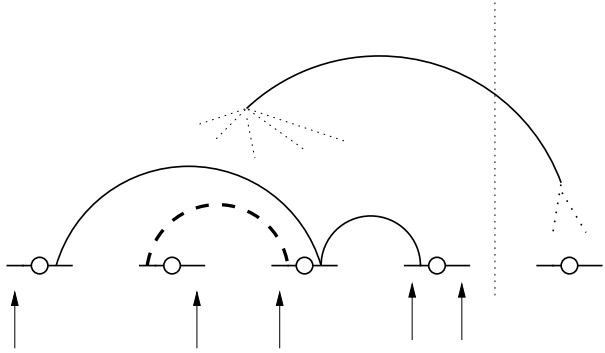}}
\caption{\footnotesize{Arc to join  $u_k$ with $P_{k-1}$. Arrows show 
possible emplacements of the left foot.}}
\end{figure}

%%%%%%%%%%%%%%%%%%%%%%%%%%%%%%%%%%%%%%%%%

Not let us pass to the case (b). In this case $P_{k-1}(\delta_{k})$
is 
splitted in two connected diagrams. Let us assume that the left element
of $u_k$ is connected by an arc with the diagram constructed on $j$ vertices.
There are $j+1$ possibilities to put the left foot of the arc. The right element of $u_k$
joined to the component of $k-1-j$ elements. There are   $k-j$ possibilities to do this. 
 The choice of the vertices to produce the component of $j$ elements
gives  ${ k-1\choose j}$ possibilities. Then we get the formula (4.3).
It is easy to see that on the way described we obtain all the diagarms of $\cd_k$.

Theorem 4.1 is proved. 
\v

{\it Proof of Proposition 4.2.}
Recurrent relation (4.4)   resembles very much  the recurrent relation determining 
the moments of the famous semi-circle distribution from  random matrix theory
 $$
m_k = v^2 \sum_{j=0}^{k-1} m_j m_{k-1-j}, \quad m_0=1
\eqno (4.13)
$$
with a parameter $v>0$ \cite{W}.
It is also known that  the numbers $m_k=m_k(v)$ are proportional to  the Catalan numbers $C_k$;
$$
m_k(v)= v^{2k} C_k= {1\over k+1} { 2k\choose k}.
\eqno (4.14)
$$
One can easily deduce from (4.13) that 
$$
m_k \le (2v)^{2k}.
$$
Comparing (4.4) with (4.13) taken for $v^2=2$, we conclude that
$$
h_k \le 8^k
\eqno (4.15)
$$
for all $k\ge 1$. This proves regularity of the generating function $h(z)$
in the domain $z\in R_{1/8}$.

Now let us show that $h(z)$ verifies equality (4.5).
It is easy to see that $h(x), -1/8<  x<1/8$ is determined by the differential equation
$$
h'(x) = {2h^2(x)\over 1-2x h(x)},\quad h(0) =1.
\eqno (4.15)
$$
Indeed, rewriting (4.4) in the form
$$
h_k =  \sum_{j=0}^{k-1} h_j \ h_{k-1-j}+
{1\over k} \sum_{j=0}^{k-1} h_j \ h_{k-1-j}, 
\eqno (4.16)
$$
we multiply both parts of (4.16) by $x^k$ and after summation over $k\ge 1$,
we get relation
$$
h(x) -1 = x h^2(x) + \int_0^x h^2(y) dy.
$$
Passing to the derivatives, we get equation (4.15).

With the help of the substitution 
 $$
x h(x) = \psi(x)
$$
we reduce (4.15) to equation
$$
\psi' = {\psi\over x(1-2\psi)}
$$
that gives
$\psi e^{-2\psi} = Cx$. 
Observing that $C=1$, we get  (4.5).

To determine the explicit form of coefficients $h_k$, we use the standard
technique of the contour integration \cite{R}.
First let us note that function $\psi (z) = z h(z)$ verifies the P\'olya equation
$$
\psi (z) = z e^{2\psi(z)}
$$
and that the inverse function $\psi^*(w) = we^{-2w}$ is regular in the vicinity of the origin.
By the Cauchy formula, we have
$$
\psi_k = {1\over 2\pi \i} \oint {\psi(z)\over z^{k+1}} dz.
$$
Changing variables by $z = \psi^*(w)$, we get $ dz = (1-2w)e^{-2w} dw$ and
find that
$$
\psi_k = {1\over 2\pi \i} \oint {1-2w\over w^k} e^{2wk} dw.
$$
Then 
$$
\psi_k = 2^{k-1} {k^{k-2}\over (k-1)!}
$$
and  (4.6) follows.
Proposition 4.2 is proved.
\v

Accepting that $d_1=1$ and introducing the exponential generating function
$$
D(\tau) = \sum_{k=1}^\infty {2^{k-1}d_k\over k!} \tau^k,
\eqno (4.17)
$$
and taking into account  (4.2),
one can write formally for the partition function (3.2) that
$$
\lim_{n\to\infty}{1\over n c} \log  \hat Z_n(\beta,{g\over c})  \simeq {e^{2g}\over 2}
D(ge^{2g}).
\eqno (4.18)
$$
The rigorous derivation of this equality will be done in separate publication. 

\section{The case of constant edge probability}

Now it is easy to prove the following statement that concerns
partition function (3.2), where $X_n$ is determined by (3.3). 
\v
\n 
{\bf Proposition 5.1}. {\it Given $k\ge 1$, there exists the limit
$$
\lim_{n\to\infty} {1\over n^2} Cum_k({g\over n} X_n) = g^k \sum_{\pi_k} W(\pi_k;p)
= g^k \sum_{\d_k \in \cd_k } w(\d_k;p),
\eqno (5.1)
$$
where $W$ and $w$ are polynomials in $p$ involving degrees $p^{k+1}, \dots, p^{2k}$
and $p= e^{-2\b'}/(1-e^{-2\b'}) $ with $\beta ' = \b -g$.}

\v
{\it Proof.}
We follow the lines of the proof of Theorem 4.1 because the formulas (4.8), (4.9) and (4.10) are still valid in the present case with obvious changes.
Regarding  (4.10), let us consider the term that corresponds to the trivial partition 
$\pi_k ^{(1)}$. It is easy to see that the formula (4.11) reads as
$$
W(\pi_k ^{(1)}) = (\E a)^k \E a^k = p^{k+1}.
$$
Clearly, relation (4.12) takes the form 
$$
\E\{\tilde Y_{T_1(\pi_k;\bar \a(\d_k))}\}\cdots
\E\{\tilde Y_{T_\s(\pi_k;\bar \a(\d_k))}\} =
{p}^{k+1+\chi},
$$
and this weight does not vanish as it was before. 
In particular, another trivial partition of the set  $U_k$ into $k$ subsets
produces the factor $p^{2k}$.
Relation (4.9) completes the proof of the first equality of (5.1). 

The second equality in (5.1) represents the another order of the summation: we fix
a partition $\pi_k$ and consider the sum over all diagrams $\d_k \in \cd_k$ 
$$
w(\d_k;p) = (-1)^{\s-1} (\s-1)! \sum_{\d_k\in \cd_k}
 \E\{\tilde Y_{T_1(\pi_k;\bar \a(\d_k))}\}\cdots
\E\{\tilde Y_{T_\s(\pi_k;\bar \a(\d_k))}\}.
\eqno (5.2).
$$
Certainly, the numbers $W(\pi_k;p)$ and $w(\d_k;p)$ are uniquely determined.
It is possible to obtain recurrent relation that determine the weight $W(\pi_k;p)$.
This recurrent relation generalize (4.3) but it is very cumbersome and complicated.
We do not present it here.

\section{Discussion and perspectives}
We have introduced the discrete analog of the matrix models with quartic potentials.
We have shown that in this approach the Erd\H os-R\'enyi ensemble naturally arises
and plays the same role as that played by GUE for the matrix models. Regarding cumulant expansion of the normalized partition function, we have shown that 
the connected diagrams on two-valent vertices replace the four-valent two-vertex maps
seen in analogous situation in matrix models. Let us discuss our results with respect to the properties of random graphs. 

Given a graph $\gamma$ with adjacency matrix $A(\gamma)$, the variable
$$
{1\over 2} X_n = {1\over 2} \sum_{i,j=1}^n (A^2)_{ij} = {1\over 2}  \sum_{i,l,j=1}^n A_{il} A_{lj}
\eqno (6.1)
$$
represents  the  number of all possible  two-step walks over $\gamma$. 

From this point of view, it is natural to ask the same question about
asymptotic behavior of the $q$-step walks and study the terms of the cumulant expansion of variable
$$
\Theta_q^{(n)}(c;g) = {1\over n } \log \E_{\b'} \left\{ e^{g X_n^{(q)}} \right\}, \quad {\hbox{with }}
\ X_n^{(q)} = \sum_{i,j=1}^n (A^q)_{ij}, \quad q\ge 2.
\eqno (6.2)
$$
It is not hard to show that in the situation of (6.2), the reasonings of Sections 3 and 4
remains true with obvious changes. The first modification is that instead of the
two-valent vertices $u_s$ we have to consider $q$-valent vertices. More precisely,
$u_s$ are represented by the linear graph with $q+1$ subvertices and $q$ edges that join them. But it is not hard to see that the number of corresponding diagrams is equivalent to the number of diagrams constructed on the set of $q$-valent vertices, or 
in other words $q$-stars with ordered (or labelled) off-spreads.
Let us count the number $d_k^{(q)}$ of connected reduced acyclic diagrams on $q$-stars.

We start with the case $q=3$.
It is not difficult to repeat the proof of Proposition 4.2 and to derive recurrent relations
for $k\ge 2$
$$
d_k^{(3)} = 3(2k -1) \, d_{k-1}^{(3)}
 + 3I_{\{k\ge 3\}}\times  \left(\sum_{j_1+j_2=k-1, j_i\ge 1} {(k-1)!\over j_1! j_2!} (2j_1+1)(2j_2+1) d^{(3)}_{j_1}
 g^{(3)}_{j_2}\right) 
 $$
$$
+I_{\{k\ge 4\}} \times \left(\sum_{j_1+j_2+j_3=k-1, j_i\ge 1} {(k-1)!\over j_1! j_2!j_3!} (2j_1+1)(2j_2+1) (2j_3+1)d^{(3)}_{j_1}
 d^{(3)}_{j_2}  d^{(3)}_{j_3}\right)
 \eqno (6.3)
 $$
 with the initial condition $d_1^{(3)} = 1$. Introducing the auxiliary numbers 
 $$
 h^{(3)}_k = (2k+1) d^{(3)}_k/k!
 $$ 
 and setting $h_0^{(3)}=1$, it is not hard to show that (6.3) is equivalent to the following  recurrent relation
 $$
 h^{(3)}_k = {2k+1\over k} \sum_{j_1+j_2+j_3 = k-1, j_i\ge 0} h^{(3)}_{j_1} h^{(3)}_{j_2} 
 h^{(3)}_{j_3}, \quad k\ge 1.
 \eqno (6.4)
 $$
 Using (6.4),  we derive 
 differential equation
 $$
 {dh^{(3)}(x)\over dx} = {3\left(h^{(3)}(x)\right)^3\over 
 1 - 6x \left(h^{(3)}(x) \right)^2}, \quad h^{(3)}(0)=1,
 $$
 where $h^{(3)}(x) = \sum_{k\ge 0} h^{(3)}_k x^k$.
 
 In the general case of $q\ge 2$, we obtain equation
 $$
 {dh^{(q)}(x)\over dx} = {q\left(h^{(q)}(x)\right)^q\over 
 1 - (q^2-q)x \left(h^{(q)}(x) \right)^{q-1}}, \quad h^{(q)}(0)=1.
 \eqno (6.5)
 $$
Substitution $\psi(x) = x \left( h^{(q)}(x)\right)^{q-1}$ leads us to equation
$$
\psi'(x) = {1\over x} \cdot { \psi(x)\over 1 - (q^2-q) \psi(x)}.
$$
Resolving it and returning back to the function $h(x) = h^{(q)}(x)$, we 
arrive at the equation that generalizes the P\'olya equation
$$
h(x) = \exp\{qxh^{q-1}(x)\}, \quad q\ge 2.
\eqno (6.6)
$$
Using again  the function $\psi$ and repeating computations of Section 4,
we obtain relations to determine the coefficients $h^{(q)}_k, k\ge    0$;
$$
\sum_{j_1+\dots+j_{q-1}=k} h_{j_1}\cdots h_{j_{q-1}} = \psi_{k+1} = 
(q^2-q)^k {(k+1)^{k-1}\over k!}, \quad h_0 = 1.
$$ 
In particular, for $q=3$, we obtain the  first values $h^{(3)}_1 = 3$, 
$h^{(3)}_2 = 45/2$, $h^{(3)}_3= 1071/6$ that correspond to the sequence
$d^{(3)}_1 = 1$, $d^{(3)}_2 = 9$, $d^{(3)}_{3} = 153$.

As we have seen, the discrete matrix model (2.5) we proposed represents an interesting
source of questions about corresponding combinatorial structures.
The limiting expressions we  compute represent the first, zero-order approximation
 to the free energy per site of this model. It could be interesting to further develop
 the diagram approach to study the next terms of the $1/n$-expansion of this free energy and justify equality (4.18). 
 A special attention is to be paid for the limiting transition $n\to\infty$, $ c=const$. One can expect that the combinatorial structure of the cumulants of $X_n$ is determined again by the diagrams of the type
 ${\cal D}_j$ but now all of the diagrams with $j=1,\dots, k$ are to be involved. This is a subject of a separate publication.

\end{document}